\documentclass[publish]{jkas} 

\def\year{2025} 
\def\volume{58} 
\def\issue{2} 
\def\beginpage{291} 
\def\received{September 8, 2025} 
\def\accepted{November 28, 2025} 
\def\published{December 1, 2025} 
\date{Received \received; Accepted \accepted; Published \published}

\usepackage{balance}
\usepackage{multirow}
\newcommand\pt[1]{\phantom{#1}}

\title{%
On-Orbit Calibration of Danuri/PolCam. I. Geometric Calibration
}

\author[1]{Kilho Baek}{0000-0002-2703-7810}
\author[2,3,$\star$]{Sungsoo S. Kim}{0000-0002-5570-2160}
\author[4]{Minsup Jeong}{0000-0002-5434-5181}
\author[4]{Young-Jun Choi}{0000-0001-6060-5851}

\affil[1]{School of Space Research, Kyung Hee University, Yongin-si, Gyeonggi-do 17104, Republic of Korea}
\affil[2]{Humanitas College, Kyung Hee University, Yongin-si, Gyeonggi-do 17104, Republic of Korea}
\affil[3]{Department of Astronomy and Space Science, Kyung Hee University, Yongin-si, Gyeonggi-do 17104, Republic of Korea}
\affil[4]{Korea Astronomy and Space Science Institute, Daejeon 34055, Republic of Korea}

\def\corrauthor{%
S. S. Kim, \email{sungsoo.kim@khu.ac.kr}
}

\def\runningauthor{%
Baek et al.
}

\def\runningtitle{%
On-Orbit Calibration of Danuri/PolCam
}

\def\keywords{%
instrumentation: polarimeters --- techniques: image processing --- Moon --- planets and satellites: surfaces
}

\def\abstracttext{%
The wide-angle Polarimetric Camera (PolCam) onboard South Korea’s first lunar orbiter, Danuri, is a pioneering instrument designed to conduct the first global polarimetric and high-phase-angle survey of the Moon. Precise geometric calibration is critical for this mission, particularly due to PolCam's highly oblique viewing geometry, which introduces significant topographic distortion. We present a comprehensive on-orbit geometric calibration that relies on 160,256 tie points derived from matching features between PolCam images and the well-orthorectified global map of the Kaguya Multiband Imager (MI). This dataset allows us to address two fundamental challenges: (1) the accurate reconstruction of the observation time for each line of an observation strip via a simple linear model, and (2) the refinement of the precise camera model, geometric model for PolCam optics. Our optimization method for these two challenges transforms the 2D image coordinates of identified features into 3D lunar coordinates and minimizes the reprojection error against the reference coordinates provided by the Kaguya MI map. From the refined observation time and camera model, we compute the precise longitude, latitude, and elevation of each pixel of an observed image. These estimated 3D coordinates are then used to generate orthorectified images, the final product of the geometric calibration. The resulting calibration achieves a geometric precision comparable to that of previous lunar orbiters and establishes the foundational framework necessary to produce geometrically-corrected data products of PolCam.
}

\begin{document}
\jkashead 


\section{Introduction}\label{sec:s1}

The wide-angle Polarimetric Camera (PolCam) onboard the Korean Pathfinder Lunar Orbiter (KPLO; also known as Danuri) is the first instrument to conduct polarimetric imaging from lunar orbit. Polarimetric observations are a powerful technique in optical remote sensing, enabling the estimation of surface properties such as median grain size \citep{Jeong2015, Shkuratov2015} and internal opacity \citep{Shkuratov2007}. Prior to PolCam, however, no such observations had been performed from lunar orbit. Although ground-based polarimetric studies exist, they are limited by constraints on phase angle, spatial resolution, and the inability to observe the lunar far-side. PolCam is expected to provide the first high-resolution, global polarimetric dataset of the Moon, offering new insights into its surface properties.

\begin{figure*}[!t]
\centering
\includegraphics[width=\linewidth]{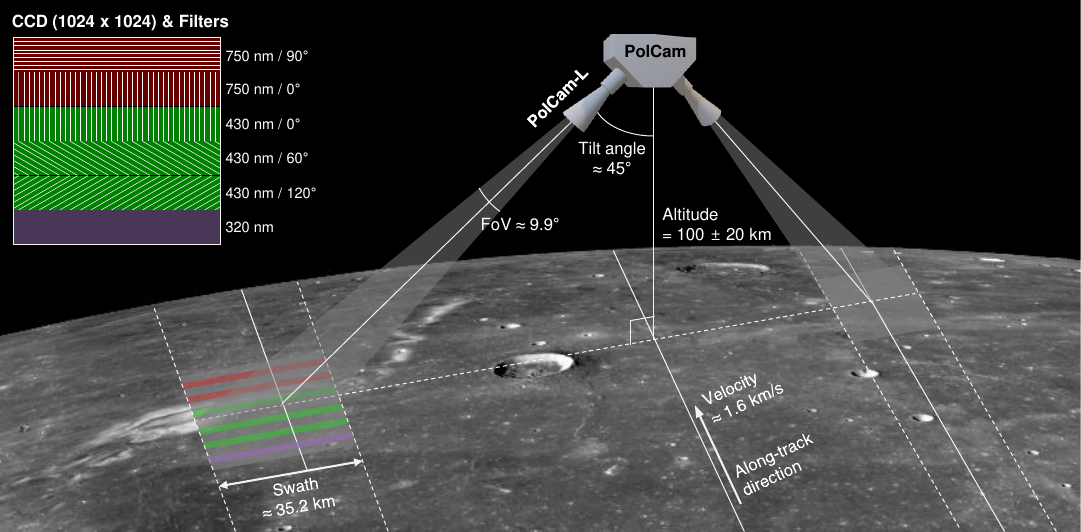}
\caption{Schematic diagram of the PolCam observation geometry in lunar orbit. The inset in the top-left corner details the central wavelength and polarizer orientation for the six channels on the $1024\times1024$ CCD. Data are acquired from only six specific lines (one line per channel) of the 1024 lines}\label{fig:f1}
\vspace{5mm}
\end{figure*}

PolCam consists of two cameras mounted at a $45^{\circ}$ tilt in the cross-track direction relative to the spacecraft’s nadir. Each camera is equipped with a $1024\times1024$ CCD (\texttt{Teledyne e2v CCD47-20}) and performs push-broom observations by collecting data from only~$6$ of the~$1024$ available horizontal lines \citep{Jeong2023}. Figure~\ref{fig:f1} provides a schematic \mbox{diagram} illustrating this observation geometry, along with the wavelength and polarizer orientation for six color/polarization angle channels. This oblique configuration was designed to enable observations at high phase angles, particularly beyond $90^{\circ}$, where the lunar polarimetric phase function shows maximum polarization near $105^{\circ}$ \citep{Sim2020}. However, this highly oblique viewing geometry also results in stronger topographic distortions in the observed images.

%
Geometric calibration is a indispensable data processing procedure that enables precise orthorectification, a process that corrects image distortions from various sources to accurately represent the undistorted (true) ground surface. This is achieved by precisely estimating the distortions introduced by the satellite's motion, the camera's optical characteristics, and the \mbox{target's} curvature and topography as light travels from the \mbox{target} surface to the CCD. When this estimation is performed using imagery acquired in orbit, it is referred to as ``on-orbit'' geometric calibration.

On-orbit geometric calibration typically involves the following steps: (1) the extraction of georeferenced features from the observed images, (2) estimation of the observation time, (3) refinement of the camera model, and (4) orthorectification. Accurate georeferences and feature extraction are essential for constructing the fundamental dataset required for optimizing the observation time and camera model. The satellite's position and attitude at the moment of observation is derived from the observation time, while the camera model describes the characteristics and distortions of the optics. A refined observation time and camera model allow for an accurate orthorectification. The parameters that define the camera model include the mounting angles, focal length, principal point, and optical distortion coefficients. Collectively, these are known as the camera parameters, and the process of refining them is called camera calibration.


Common types of georeferences include digital terrain model (DTM), ground control points (GCPs), and tie points. GCPs, such as lunar laser ranging retroreflectors \cite[LRRRs;][]{Kokurin1972, Harvey2007, Murphy2007, Murphy2011, Battat2009}, are ground truth features with well-known coordinates used to georeference an image for absolute orientation. In contrast, tie points are features with unknown coordinates that are visible in overlapping images; they link the images together to establish relative orientation.


High-precision geometric calibration, which is essential for achieving sub-pixel geolocation accuracy, ensures channel alignment for multi-spectral instruments and enables \mbox{consistent} comparison with datasets from previous missions. In the case of PolCam, it is especially critical due to its off-nadir viewing geometry, which introduces more severe topography-induced distortions than nadir-pointing systems. Furthermore, precise inter-channel registration is even more vital than for conventional optical observations, as it directly impacts the accuracy of the derived degree of polarization.

Past lunar orbiters, including Clementine’s Ultraviolet/Visible (UVVIS) camera \citep{Kordas1995, Nozette1994}, LRO’s LROC Narrow Angle Camera (NAC) and Wide Angle Camera (WAC) \citep{Robinson2010}, Kaguya’s Multiband Imager \citep[MI;][]{Ohtake2008}, and stereo camera of Chang’e-1 and~2 \citep{Di2012}, achieved geometric calibration through a combination of pre-flight laboratory measurements and on-orbit refinements. Typically, intrinsic parameters were first determined from optical bench tests, followed by on-orbit corrections using control points or tie points matching, bundle adjustment, and alignment with digital elevation models (DEMs).

For example, the LROC NAC obtained highly accurate camera parameters before launch and used LRRRs on the surface as ground control points to estimate internal timing. For the LROC WAC, camera parameters were estimated by co-registering WAC images with simultaneously acquired NAC images \citep{Speyerer2016}. Kaguya’s MI camera calibration combined star field observations with nadir stereo imaging \citep{Ohtake2010}. Chang’e-2 mission applied photogrammetric block adjustment to overlapping stereo images, and proposed a self-calibration bundle adjustment method to address inconsistencies between forward- and backward-looking images \citep{Di2014}.

Most Earth-observing satellites perform geometric calibration using terrestrial landmarks as GCPs \citep{Lee2004, Takaku2009, Storey2014, Wang2014, Wang2017, Zhang2023}. Satellites observing regions without suitable landmarks often perform self-calibration using tie points extracted from multiple overlapping images \citep{Jiang2018, Wang2018, Yang2020}. For planetary orbiters, which also lack defined landmarks, calibration is typically achieved either through self-calibration using tie points from two cameras \citep{Ohtake2010, Di2014, Speyerer2016, Denevi2018} or by matching a DTM generated from stereo imagery with one DTM measured by a laser altimeter \citep{Li2023}. Performing geometric calibration using tie points derived from another mission's calibrated dataset is uncommon \citep{Speyerer2016clem, Speyerer2023, Dong2019}. However, due to insufficient pre-launch ground calibration and the absence of a stereo camera for self-calibration, PolCam adopted the method, extracting tie points using the Kaguya MI map.

Unfortunately, PolCam’s camera model was not accurately measured during pre-launch ground tests, and the precise observation times are not recorded in the telemetry headers---only the storage time of compressed 64-line image packets is available. In addition, the spatial resolution of PolCam is insufficient to use LRRRs as ground control points, as done in the case of LROC NAC. Consequently, we were faced with the challenge of performing geometric calibration using only on-orbit observation data.

In the present study, we performed feature detection and matching between PolCam imagery and the Kaguya MI dataset to construct a network (i.e., a large bundle) of tie points. Using this network, we performed a high-precision geometric calibration by successfully estimating both the observation time for each horizontal line of an observation strip and the camera model of PolCam. Section~\ref{sec:s2} describes the reference datasets, including the lunar global map for ground coordinates and the ancillary data used to reconstruct the observation geometry. Section~\ref{sec:s3} outlines the methodology for extracting tie points and performing the geometric calibration. Sections~\ref{sec:s4} and~\ref{sec:s5} present the results of the observation timing estimation and camera model refinement, respectively. Section~\ref{sec:s6} showcases the geometrically-corrected PolCam images. Section~\ref{sec:s7} concludes with a summary and discussion.

In addition, we are preparing a series of papers related to the on-orbit calibration of PolCam. The present study is the first, followed by a second paper on smear artifact correction and a third on photometric calibration. PolCam data suffer from a significant amount of smear artifacts, which critically impact the calculation of the degree of polarization, making their removal crucial for polarization accuracy. Although photometric calibration is not essential for PolCam's primary objective of generating polarization map, it will constitute a meaningful study, as its observations at phase angles greater than 90 degrees are unprecedented for a lunar orbiter.

\begin{figure*}[!t]
\centering
\includegraphics[width=\linewidth]{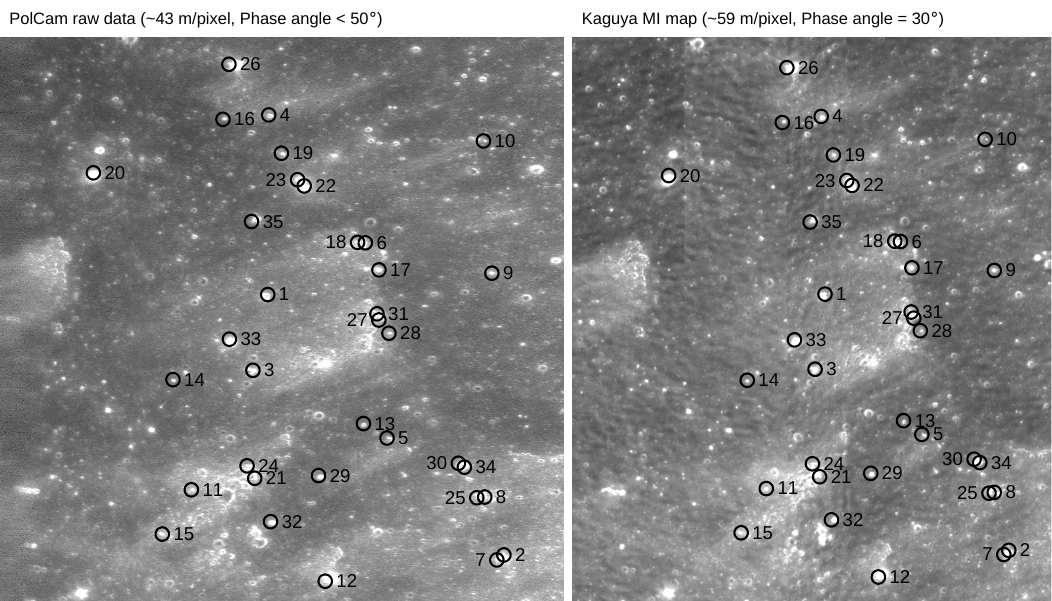}
\caption{A subset of tie points extracted by feature detection and matching techniques. (Left) A sub-region of a PolCam raw image. (Right) The corresponding sub-region of the Kaguya MI map. Circular symbols in both images mark the locations of detected tie points, with numeric labels identifying the matched pairs.}\label{fig:f2}
\vspace{2mm}
\end{figure*}

\newpage
\section{Data Preparation}\label{sec:s2}

To establish our ground reference, we used the topographically-corrected reflectance map\footnote{\url{https://planetarymaps.usgs.gov/mosaic/Lunar_MI_multispectral_maps/index.html}} from the Kaguya MI \citep{Lemelin2019, Ohtake2013} and the improved lunar digital elevation model \citep[known as SLDEM;][]{Barker2016}\footnote{\url{https://astrogeology.usgs.gov/search/map/moon_lro_lola_selene_kaguya_tc_dem_merge_60n60s_59m}}, which combines data from the LRO Lunar Orbiter Laser \mbox{Altimeter} (LOLA) and the Kaguya Terrain Camera (TC). The Kaguya MI map, covering latitudes up to $\pm65^{\circ}$ with a spatial resolution of $512$~pixels per degree (PPD), is well-suited for comparison with PolCam’s data ($\pm70^{\circ}$ latitude coverage and $\sim$$700$~PPD resolution). However, a key consideration is that the Kaguya MI map is normalized to a $30^{\circ}$ phase angle $(\text{incidence} = 30^{\circ}, \text{reflectance} = 0^{\circ})$. To ensure comparable illumination conditions for reliable feature matching, we selected a specific subset of 40~PolCam orbits (out of more than~8,012 available) acquired at phase angles $50^{\circ}$ $(\text{incidence} < 5^{\circ}, \text{reflectance} \approx 45^{\circ})$.

The observation geometry was determined using the Spacecraft, Planet, Instrument, C-matrix, Events (SPICE) kernels and Toolkit \citep{Acton1996}. Developed by NASA’s Navigation and Ancillary Information Facility (NAIF), the SPICE system provides the essential tools for computing the observation geometry, orientation, and timing of spacecraft and their payloads. Spacecraft Position Kernels (SPKs) provided the spacecraft's location and the mounting position of PolCam, while C-Matrix Kernels (CKs) provided the spacecraft's attitude. The Frame Kernel (FK) and Instrument Kernel (IK) were used to define the instrument's extrinsic and intrinsic parameters, respectively. The SPICE kernels for the Danuri mission are provided by the Korea Aerospace Research Institute (KARI) through the KPLO Data System\footnote{\url{https://www.kari.re.kr/kpds/}} (KPDS).

Since Danuri does not carry its own on-board altimeter, the viewing geometry must be determined using a digital shape model derived from previous missions. The $100~\mathrm{m/pixel}$ resolution Digital Shape Kernel (DSK) provided by European Space Agency (ESA) through an online archive\footnote{\url{https://spiftp.esac.esa.int/data/SPICE/esa_generic/kernels/dsk/tiled_dsk/}} was too coarse relative to PolCam's spatial resolution and therefore inadequate for high-precision geometric calibration. To overcome this limitation, we selected the highest-resolution model available, the aforementioned SLDEM, and converted it into the DSK format compatible with the SPICE toolkit using the \texttt{mkdsk} SPICE utility\footnote{\url{https://naif.jpl.nasa.gov/pub/naif/toolkit_docs/C/ug/mkdsk.html}}.

\section{Method}\label{sec:s3}

As mentioned in Section~\ref{sec:s1}, high-precision geometric calibration involves accurately characterizing the distortions light undergoes as it travels from the target surface to the CCD. The most critical components are precisely estimating the observation time, which defines the satellite's position and attitude, and refining the camera model. This requires extracting and matching features with known coordinates from the observed imagery.

This section introduces the method for extracting matched pairs, referred to as ``tie points'', from PolCam images and the Kaguya MI map. We then describe the method for optimizing the observation time and camera parameters using these tie points. These features are refered to as tie points rather than ground control points because, although their coordinates are derived from the Kaguya MI map, they are estimated locations with inherent uncertainties, not absolute ground truth.


To identify and match the corresponding features between PolCam imagery and the Kaguya MI map, we employed the Oriented FAST and Rotated BRIEF (ORB) algorithm \citep{Rublee2011}, which is available in the OpenCV library. This algorithm combines an oriented Features from Accelerated Segment Test (FAST) detector with a rotated Binary Robust Independent Elementary Features (BRIEF) descriptor. OpenCV is a widely used open-source computer vision library for Python that provides efficient implementations of various algorithms, including object tracking, filtering, and facial recognition, under a permissive license. The oriented FAST algorithm efficiently detects corner features, while the rotation-invariant BRIEF descriptor generates compact binary signatures for each key point, enabling robust correspondence even under image rotation. By integrating these techniques, we achieve fast and reliable tie points, which is essential for accurate geometric calibration.

\begin{figure*}[!t]
\centering
\includegraphics[width=\linewidth]{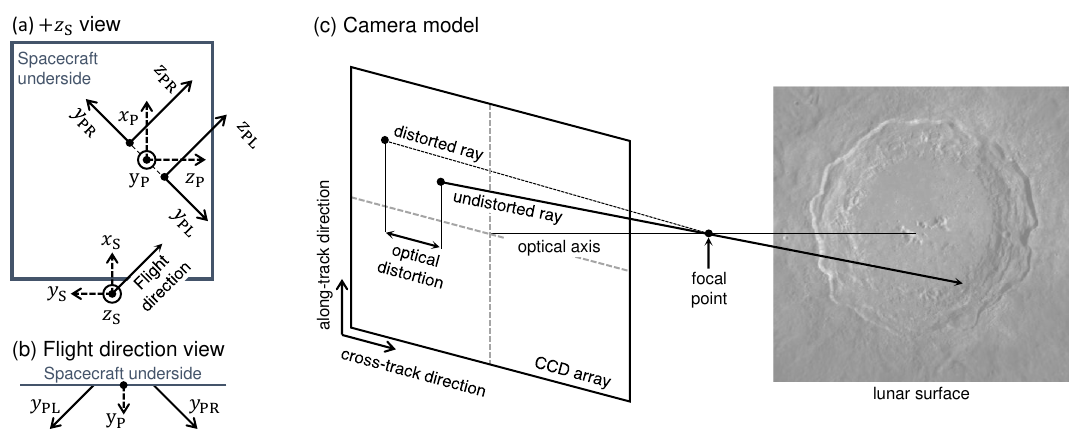}
\caption{Schematic of the PolCam instrument frame and its camera model. (a) Frame orientation of PolCam $(x_\mathrm{P}, y_\mathrm{P}, z_\mathrm{P})$, PolCam-L $(x_\mathrm{PL}, y_\mathrm{PL}, z_\mathrm{PL})$, and PolCam-R $(x_\mathrm{PR}, y_\mathrm{PR}, z_\mathrm{PR})$ relative to the spacecraft frame $(x_\mathrm{S}, y_\mathrm{S}, z_\mathrm{S})$. (b) A view of the PolCam-L $(y_\mathrm{PL})$ and PolCam-R $(y_\mathrm{PR})$ cameras along the flight direction. (c) The simplified pinhole camera model, which defines the ideal projection geometry. Optical distortion presents a misalignment between this ideal projection and the actual observed pixel location.}\label{fig:f3}
\vspace{5mm}
\end{figure*}

Figure~\ref{fig:f2} shows a subset of the tie points extracted by comparing PolCam image with the Kaguya MI map. The detected features generally correspond to small and bright points that stand out against their background terrain. As indicated by the circular symbols and corresponding numeric labels, these points were successfully matched between the two images. A total of 160,256 tie points was identified across all PolCam channels from 40~observation strips, with 18,725, 33,883, 34,996, 34,737, 22,290, and 15,625 pairs from channels~1 through~6, respectively. For each tie point, the 2D image pixel coordinates were obtained from the PolCam images, while the corresponding geographic latitude and longitude were derived from the Kaguya MI map. To construct the 3D coordinates on the lunar surface, the radius for each point was then retrieved from the SLDEM using these geographic coordinates.

\begin{table}[!t]
\centering\setlength{\tabcolsep}{5pt}
\caption{Design values of camera parameters and CCD specifications for the left-side camera (PolCam-L).}\label{tab:t1}
\begin{tabular}{cc}
\toprule
 & Specifications \\
\midrule
Mounting Angles (yaw, tilt, roll) & $(45^{\circ}, 45^{\circ}, 0^{\circ})$ \\
Field of View & $9.9^{\circ}$ \\
Focal Length ($f$) & $76.39\mathrm{mm}$ \\
Principal Point $(x_\mathrm{c}, y_\mathrm{c})$ & $(512.5, 512.5)$ \\
CCD Size & $1024\times1024$ \\
CCD Pixel Pitch & $13~\mu\mathrm{m}$ \\
\multirow{3}{*}{\makecell{Readout\\Lines}} & Ch.~1: \pt{0}48, Ch.~2: 180, \\
 & Ch.~3: 368, Ch.~4: 564, \\
 & Ch.~5: 724, Ch.~6: 884\pt{,} \\
\bottomrule
\end{tabular}
\vspace{2mm}
\end{table}

The 3D coordinates derived from the Kaguya MI map and SLDEM are defined in the Mean Earth (ME) body-fixed reference system \citep[$\mathrm{MOON}\_\mathrm{ME}$;][]{Seidelmann2007, LRO2009}. To project these ground coordinates onto the corresponding PolCam image pixels, a series of frame transformations was performed in the following sequence: $\mathrm{MOON}\_\mathrm{ME}$ $\rightarrow$ spacecraft $\rightarrow$ PolCam $\rightarrow$ PolCam-L $\rightarrow$ image plane. This transformation process utilizes PolCam's extrinsic and intrinsic parameters and the design values for which are listed in Table~\ref{tab:t1}. Furthermore, this study considered only the left-side camera (PolCam-L frame), which was actively used for \mbox{observation}.\footnote{The right-side camera is not functioning as intended, but this issue does not compromise the planned scientific objectives. The left-side camera alone has successfully provided global lunar coverage within a phase angle range of $45^{\circ}$ to $135^{\circ}$.} Ultimately, each tie point was represented by its 2D coordinates on the image plane, relative to the principal point, and expressed in units of pixels for analytical convenience.

The coordinate systems for the spacecraft and the PolCam instrument are illustrated in Figure~\ref{fig:f3}. The spacecraft frame (S) is defined with its $+x_\mathrm{S}$ axis pointing in the forward direction of the spacecraft and its $+z_\mathrm{S}$ axis pointing toward nadir. The PolCam frame (P) is fixed to the spacecraft's nadir deck, rotated $+90^{\circ}$ about the $+x_\mathrm{S}$ axis. This rotation aligns the $+y_\mathrm{P}$ axis with the spacecraft's nadir-pointing $+z_\mathrm{S}$ direction. The individual camera frames are defined relative to the PolCam frame. By flight model design, the PolCam-L camera frame (PL) is oriented by a $+45^{\circ}$ yaw rotation (about $+y_\mathrm{P}$) followed by a $+45^{\circ}$ tilt rotation (about the new x-axis), as depicted in Figure~\ref{fig:f3}(a). The camera's boresight is aligned with its $+y_\mathrm{PL}$ axis. A view along the flight (or along-track) direction in Figure~\ref{fig:f3}(b) shows the boresights for both the left and right cameras, showing the instrument's design for off-nadir observations at a $45^{\circ}$ tilt angle.

\begin{figure}[!t]
\centering
\includegraphics[width=\linewidth]{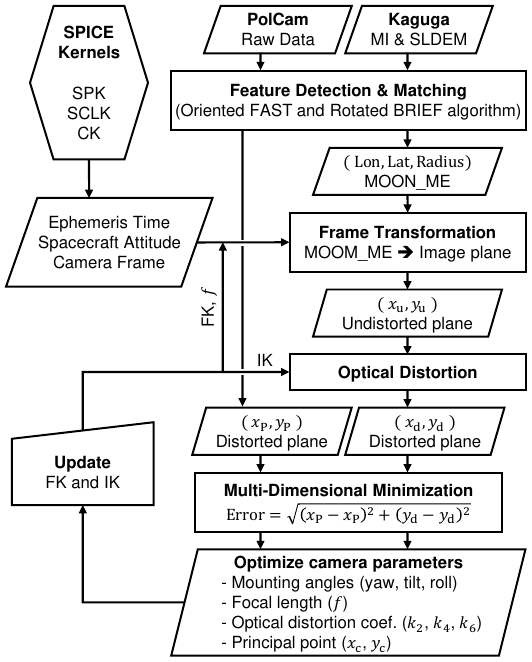}
\caption{Flowchart of the camera calibration process.}\label{fig:f4}
\vspace{2mm}
\end{figure}

The pre-flight design values for the extrinsic and intrinsic parameters (Table~\ref{tab:t1}) are insufficient to meet the requirements for precise geometric calibration. In other words, the computed pixel coordinates for reference ground points do not match their observed locations in the PolCam image [Figure~\ref{fig:f3}(c)], largely due to unmodeled optical distortion and mounting uncertainties. This misalignment means that the true viewing geometry for a given observation cannot be accurately determined using the design values. Therefore, we need to determine the true in-flight values of these parameters via on-orbit geometric calibration. This calibration process is illustrated in Figure~\ref{fig:f4} and consists of three sequential steps: (1) constructing a network of tie points, (2) performing frame transformations, and (3) optimizing the camera parameters.

The parameters are broadly divided into two categories: Extrinsic parameters define the camera's position and orientation relative to the spacecraft body; this orientation is typically described by yaw, pitch, and roll rotation angles. Intrinsic parameters, in contrast, describe the inherent properties of the camera's optical system; These include the effective focal length, principal point, and optical distortion coefficients.

\begin{figure*}[!t]
\centering
\includegraphics[width=\linewidth]{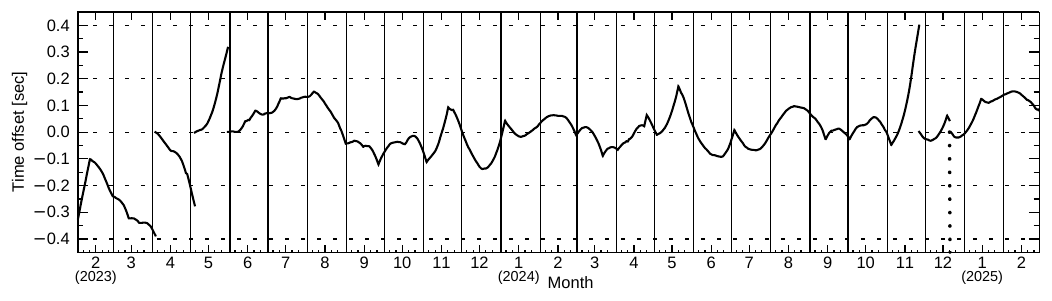}
\caption{Time offset between the Danuri spacecraft clock and ephemeris time from Feburary 2023 to February 2025.}\label{fig:f5}
\vspace{3mm}
\end{figure*}

To project a ground point onto the PolCam detector, we first define the vector pointing from the PolCam-L camera to the ground point:
\begin{equation}\label{eq:e1}
\hat{r}_\mathrm{PM} = \frac{r_\mathrm{M} - r_\mathrm{P}}{|r_\mathrm{M} - r_\mathrm{P}|}
\end{equation}
where $r_\mathrm{M}$ is the position vector of a point on the lunar surface, $r_\mathrm{P}$ is the position vector of the PolCam-L frame’s center, and $\hat{r}_\mathrm{PM}$ is the unit vector pointing from the PolCam-L to that surface point. All vectors are initially defined in the $\mathrm{MOON}\_\mathrm{ME}$ frame.

Next, using the transformations defined in the SPICE Camera and Frame Kernels, this unit vector is converted from the $\mathrm{MOON}\_\mathrm{ME}$ to the PolCam-L frame:
\begin{equation}\label{eq:e2}
[x, y, z]^\mathrm{T} = \mathrm{M}_3 \mathrm{M}_2 \mathrm{M}_1  \hat{r}_\mathrm{PM}^\mathrm{T}
\end{equation}
where $[x, y, z]$ is the unit vector viewing a ground point in the PolCam-L frame. $\mathrm{M}_1$, $\mathrm{M}_2$, and $\mathrm{M}_3$ are the frame transformation matrices for the $\mathrm{MOON}\_\mathrm{ME}$ $\rightarrow$ spacecraft, spacecraft $\rightarrow$ PolCam, and PolCam $\rightarrow$ PolCam-L transformations, respectively. This 3D vector in the instrument frame is then projected onto the 2D image plane to derive the ideal (undistorted) pixel coordinates using the effective focal length $(f)$ [see Figures~\ref{fig:f3}(a) and~\ref{fig:f3}(c)]:
\begin{equation}\label{eq:e3}
x_\mathrm{u} = f \times \frac{x}{z}
y_\mathrm{u} = f \times \frac{y}{z}
\end{equation}

Finally, the radial distortion model is applied to calculate the distorted pixel coordinates $(x_\mathrm{d}, y_\mathrm{d})$ that correspond to pixel of tie-point on the PolCam image:
\begin{align}\label{eq:e4}
\begin{split}
x_\mathrm{d} &= x_\mathrm{c}+(x_\mathrm{u}-x_\mathrm{c}) \times (1 + k_2 r^2 + k_4 r^4 + k_6 r^6) \\
y_\mathrm{d} &= \:\!y_\mathrm{c}+(\:\!y_\mathrm{u}-\;\!y_\mathrm{c}) \times (1 + k_2 r^2 + k_4 r^4 + k_6 r^6)
\end{split}
\end{align}
where $(x_\mathrm{u}, y_\mathrm{u})$ are the undistorted image coordinates, $(x_\mathrm{c}, y_\mathrm{c})$ is the principal point, and the terms $k_2$, $k_4$, $k_6$ are the radial optical distortion coefficients.

The camera parameters are then optimized by minimizing the reprojection error between the observed pixel coordinates of a tie point on the PolCam image $(x_\mathrm{P},y_\mathrm{P})$ and the computed distorted coordinates $(x_\mathrm{d},y_\mathrm{d})$ of the corresponding ground point. This error is defined as the Euclidean distance:
\begin{equation}\label{eq:e5}
\mathrm{Error} = \sqrt{(x_\mathrm{P}-x_\mathrm{d})^2 + (y_\mathrm{P}-y_\mathrm{d})^2 }
\end{equation}

The observation timing optimization was performed using only those tie points located near the CCD center, where distortions attributable to the camera parameters are negligible. The fundamental calculation flow is identical to the flowchart in Figure~\ref{fig:f4}. This process involved the simultaneous optimization of the observation time and the extrinsic parameters (yaw, tilt, roll angles), while the intrinsic parameters (focal length, principal point, and optical distortion coefficients) were excluded.

After iteratively refining the observation timing and camera parameters, we compute the precise 3D coordinates for each pixel using the \texttt{sincpt}\footnote{\url{https://naif.jpl.nasa.gov/pub/naif/toolkit_docs/C/cspice/sincpt_c.html}} function within the SPICE Toolkit. The \texttt{sincpt} function determines the 3D coordinates of the intersection point where a given pixel’s line-of-sight vector intercepts the surface of a target body. When combined with a DSK, it can precisely model topographic distortions caused by the lunar terrain. For this purpose, we utilized the highest-resolution available elevation model, SLDEM.

We employed \texttt{patch orthorectification}, a computationally efficient technique widely adopted in planetary mission, such as the Integrated Software for Imagers and Spectrometers (ISIS) system of United States Geological Survey (USGS) \citep{Anderson2013}. This approach avoids the intensive processing of rigorous pixel-by-pixel orthorectification by dividing a large scene into smaller, manageable patches, or tiles \citep{Chen2004}. Within each patch, a simpler, approximate mapping function is used to model the geometric distortion, which is itself derived from the rigorous sensor model applied only at a few anchor points. Pixels within each patch are resampled using bilinear interpolation, which is sufficient as the native resolution of the PolCam raw data ($\sim$$700$~PPD) exceeds that of the reference map ($\sim$$512$~PPD). The primary advantages of this method are a significant reduction in computational load and processing time, making it highly effective for handling the large-scale data volumes common to orbital missions.

\section{Observation Timing}\label{sec:s4}

PolCam does not have an independent internal clock; instead, it synchronizes its timing with the Danuri spacecraft clock each time it is powered on for an observation. The spacecraft clock is the on-board timekeeping mechanism that governs most spacecraft and payload events. Its accuracy is particularly critical for PolCam, as a precise estimation of the imaging start time is essential for geometric calibration. This on-board clock naturally exhibits an offset and drift relative to standard time systems like Ephemeris Time (ET) and Coordinated Universal Time (UTC). The Spacecraft CLock Kernel (SCLK), a component of the SPICE system, provides the transformation data required to correct for these discrepancies and precisely transform the spacecraft clock time to the standard time systems.

\begin{figure}[!t]
\centering
\includegraphics[width=\linewidth]{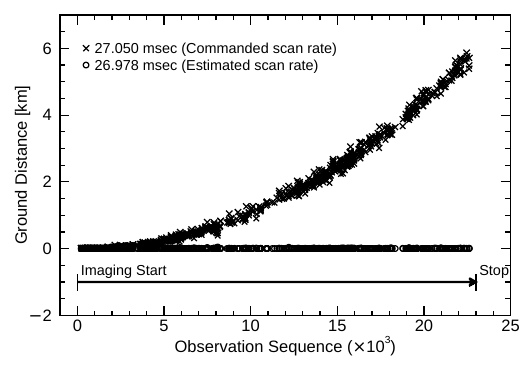}
\caption{Residual ground distance of the tie points extracted from PolCam and Kaguya MI imagery as a function of the observation sequence. The cross symbols show the growing ground distance resulting from the original commanded scan rate, while the circle symbols show the minimized ground distance achieved with the optimal scan rate estimated in this study.}\label{fig:f6}
\vspace{2mm}
\end{figure}

As mentioned in Section~\ref{sec:s1}, the PolCam telemetry headers do not record the precise observation time for each horizontal line of an observation strip. Instead, they only contain the storage time for each 64-line compressed image packet. Given this limitation, the true imaging start time for each observation must be estimated by relying on a combination of three sources: the imaging start time specified in the mission request command, the timing from the on-board spacecraft clock, and the high-precision corrections provided by the SCLK.

Figure~\ref{fig:f5} shows the time offset between the Danuri spacecraft clock and ET from February 2023 to February 2025. According to the Danuri SCLK provided by KPDS/KARI, the spacecraft clock exhibits a maximum offset of $\pm0.4~\mathrm{sec}$. Since Danuri orbits at an altitude of $100 \pm 20~\mathrm{km}$ with a velocity of $\sim$$1.6~\mathrm{km\,s^{-1}}$, this timing error translates to an along-track pointing error of $\sim$$640~\mathrm{m}$. For PolCam, with its resolution of $\sim$$700~\mathrm{PPD}$, this corresponds to an error of $\sim$$15~\mathrm{pixels}$ in the imagery.

Using SCLK data, we implemented an automated method to estimate the imaging start time for each PolCam observation. This approach achieved sub-pixel accuracy for the vast majority of the more than 8,012~orbits analyzed. However, the automated process failed for $<$$1\%$ of the cases ($56$~orbits). For these orbits, the imaging start time was re-estimated manually by visually comparing the PolCam image against the predicted viewing geometry. We speculate that these exceptions were caused by the loss of corrupted telemetry packets acquired at the very beginning of an observation sequence.

Even after verifying the imaging start time, a discrepancy between the PolCam images and the predicted viewing geometry emerged towards the end of each observation sequence. We hypothesized that this was caused by a deviation in the actual scan rate---the time interval between frames---from the value specified in the mission request command. We estimated the optimal scan rate using the network of tie points derived from comparing PolCam and Kaguya MI imagery (as described in Section~\ref{sec:s3}). Because this analysis must precede the camera calibration, we exclusively used tie points located near the center of the CCD (within $\pm64~\mathrm{pixels}$), where the effects of optical distortion are negligible.

\begin{table*}[!t]
\centering\setlength{\tabcolsep}{6pt}
\caption{Estimated camera parameters for the left-side camera (PolCam-L).}\label{tab:t2}
\begin{tabular}{cccccccc}
\toprule
 & Extrinsic Parameters & \multicolumn{6}{c}{Intrinsic Parameters} \\
\cmidrule(r){2-2}\cmidrule{3-8}
 & Mounting Angles & Focal Length & \multicolumn{2}{c}{Principal Point} & \multicolumn{3}{c}{Optical Distortion Coefficients} \\
\cmidrule(r){2-2}\cmidrule(r){3-3}\cmidrule(r){4-5}\cmidrule{6-8}
 & $(\psi, \theta, \phi)~[^{\circ}]$ & $f~[\mathrm{mm}]$ & $x_\mathrm{c}$ &  $y_\mathrm{c}$ & $k_2$ &  $k_4$ & $k_6$ \\
\midrule
Ch.~1 & \multirow{6}{*}{\makecell{Yaw $(\psi)$: 45.055\\Pitch $(\theta)$: 44.774\\Roll $(\phi)$: \pt{0}0.004}} & $76.627830$ & $512.689$ & $513.232$ &      $-1.055\times10^{-3}$ & $\pt{+}2.592\times10^{-5}$ &      $-1.515\times10^{-7}$ \\
Ch.~2 & & $75.045065$ & $512.751$ & $513.004$ &      $-2.479\times10^{-4}$ & $\pt{+}1.548\times10^{-5}$ &      $-1.363\times10^{-7}$ \\
Ch.~3 & & $74.706424$ & $512.668$ & $512.865$ & $\pt{+}1.511\times10^{-4}$ & $\pt{+}3.399\times10^{-6}$ &      $-1.850\times10^{-8}$ \\
Ch.~4 & & $74.728619$ & $512.424$ & $513.819$ & $\pt{+}2.777\times10^{-4}$ &      $-2.672\times10^{-6}$ & $\pt{+}3.567\times10^{-8}$ \\
Ch.~5 & & $74.870414$ & $512.435$ & $513.895$ & $\pt{+}2.197\times10^{-5}$ & $\pt{+}7.655\times10^{-6}$ &      $-9.458\times10^{-8}$ \\
Ch.~6 & & $76.820351$ & $512.351$ & $511.862$ &      $-1.953\times10^{-3}$ & $\pt{+}5.734\times10^{-5}$ &      $-5.048\times10^{-7}$ \\
\bottomrule
\end{tabular}
\vspace{2mm}
\end{table*}

The scan rate specified in the mission command was 27.050~$\mathrm{msec}$. However, when this value is used to calculate the ground distance between sequential tie points, the resulting error progressively increases throughout the observation sequence (as shown by the cross symbols in Figure~\ref{fig:f6}). We therefore determined the optimal scan rate that minimizes this residual ground distance, arriving at a value of 26.978~$\mathrm{msec}$, slightly shorter than commanded.

The observation time for each line of PolCam data can be expressed with the following simple linear equation:
\begin{equation}\label{eq:e6}
\mathrm{ET}_N  = \mathrm{IST} + \mathrm{SR} \times N
\end{equation}
where $\mathrm{ET}_N$ is the ephemeris time of the $N$th acquired line, $\mathrm{IST}$ is the SCLK-corrected imaging start time, and $\mathrm{SR}$ is the estimated scan rate $(= 0.026978~\mathrm{sec})$. This result was found to be consistent across the entire PolCam observation dataset.

\newpage
\section{Camera Calibration}\label{sec:s5}

Camera calibration is the fundamental process of determining the geometric and optical characteristics of a camera, which are collectively known as its camera parameters. These parameters are defined as follows: the extrinsic parameters [yaw $(\psi)$, pitch $(\theta)$, and roll $(\phi)$ angles] and the intrinsic parameters [focal length $(f)$, principal point $(x_c, y_c)$, and optical distortion coefficients $(k_2, k_4, k_6)$].

The camera calibration of PolCam was performed in a two-step process. In the first step, the extrinsic parameters (Table~\ref{tab:t2}) and a set of shared intrinsic parameters [Equation~\eqref{eq:e7}] were simultaneously estimated using the entire network of tie points from all channels. Here, the term `shared' signifies that a single set of intrinsic parameters was assumed to be common across all channels for this initial estimation. However, when a geometric calibration is performed using only these shared parameters, a residual inter-channel misalignment of 3--5 pixels persists at the image edges. This discrepancy arises because each channel has a unique combination of color and polarizing filters, with the polarizing filters being physically distinct components. This indicates that unique intrinsic parameters must be determined for each channel. Therefore, in the second step, these shared parameters served as an initial condition to estimate the unique intrinsic parameters for each individual channel (Table~\ref{tab:t2}).

\subsection{Extrinsic Parameters}\label{sec:s51}

Extrinsic parameters in camera calibration typically define an instrument's position and orientation relative to a parent frame. For PolCam, the extrinsic position was precisely measured after its integration with the spacecraft body. The center of the PolCam frame is located at $(130.1700, -13.2800, 44.6300)~\mathrm{cm}$ relative to the spacecraft frame, and the center of the PolCam-L frame is subsequently defined at $(-3.1028, -1.8187, 5.0660)~\mathrm{cm}$ relative to the PolCam frame (see Figure~\ref{fig:f3}). However, the extrinsic orientation was not measured for the flight model pre-launch. Consequently, it could only be determined by relying exclusively on on-orbit observational data.

The extrinsic parameters define the camera's mounting orientation on the spacecraft body and consist of yaw, pitch, and roll angles. For PolCam, the pitch angle is equivalent to the camera's tilt angle. The design values for the (yaw, tilt, roll) angles of the PolCam-L frame were $(45^{\circ}, 45^{\circ}, 0^{\circ})$, with a specified tolerance of $\pm1^{\circ}$. However, a non-negligible discrepancy was found between the observed images and the viewing geometry calculated using these design values. Even without accounting for intrinsic errors of the optical system, a noticeable shift in the image center was observed in the cross-track direction. This indicated a need to refine the extrinsic parameters. Therefore, we estimated these angles using the network of tie points from all channels (as described in Section~\ref{sec:s3}), with the design values in Table~\ref{tab:t1} serving as the initial condition for the calibration. Our estimated mounting angles are $(45.055^{\circ}, 44.774^{\circ}, 0.004^{\circ})$, which fall well within the specified tolerance of $\pm1^{\circ}$.

For a camera observing from a $100~\mathrm{km}$ altitude, a pointing tolerance of $\pm1^{\circ}$ is not particularly stringent. This level of uncertainty corresponds to a geolocation error of approximately $\pm1.7~\mathrm{km}$ on the lunar surface for a nadir observation, and about $\pm3.5~\mathrm{km}$ for a $45^{\circ}$ oblique observation. For PolCam, with a spatial resolution of $\sim$$43~\mathrm{m/pixel}$, a pointing accuracy of within $\pm0.001^{\circ}$ is required to maintain sub-pixel geolocation precision. Therefore, expressing the mounting angles to 3 decimal places, as shown in Table~\ref{tab:t2}, is sufficient to demonstrate the high accuracy of geometric calibration. For operational use in the official Frame Kernel, the estimated values were recorded to 6 decimal places for maximum precision.

\begin{figure*}[!t]
\centering
\includegraphics[width=\linewidth]{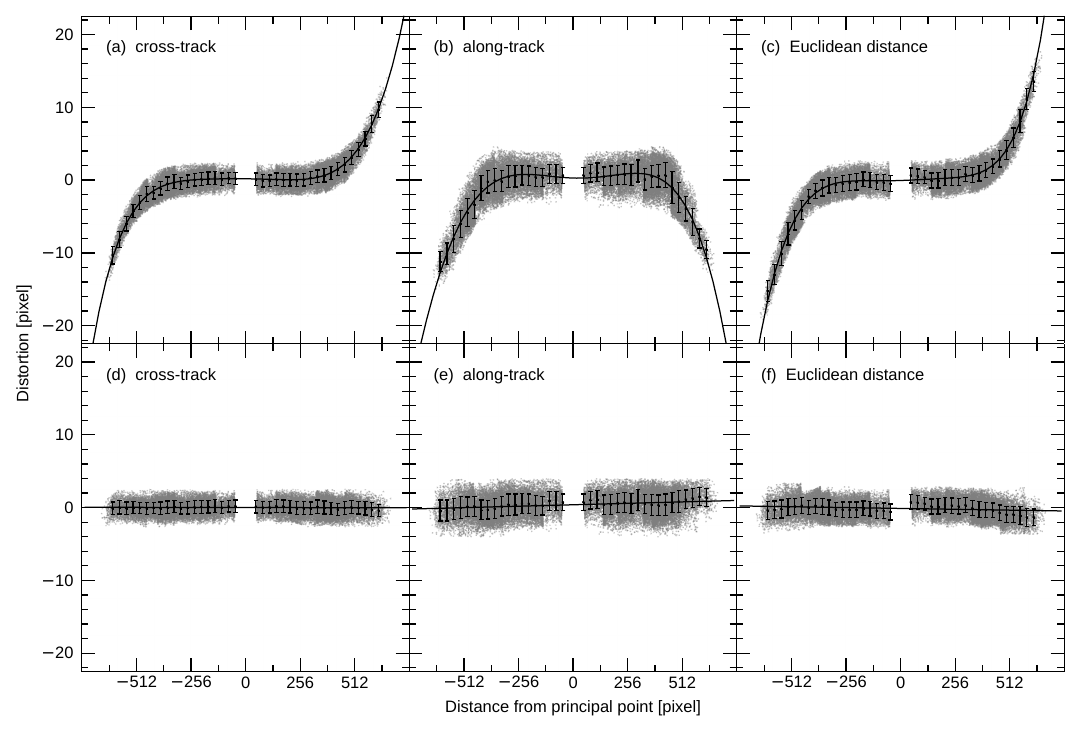}
\caption{Distortion between the tie points on the PolCam-L image plane before [top row, (a)--(c)] and after [bottom row, (d)--(f)] camera calibration. The columns, from left to right, depict the distortion in the cross-track direction, the along-track direction, and the Euclidean distance, respectively. All channels are plotted in each panel.}\label{fig:f7}
\vspace{2mm}
\end{figure*}

\subsection{Intrinsic Parameters}\label{sec:s52}

Accurate estimation of the intrinsic parameters is essential for correcting the lens distortion inherent to any optical system. These parameters are composed of the effective focal length, principal point, and optical distortion coefficients. Pre-launch ground tests of the PolCam flight model indicated a focal length of $76.39~\mathrm{mm}$, with the principal point offset and optical distortion measured to be at ideal, sub-pixel levels. However, as was the case with the extrinsic parameters, on-orbit data revealed significant discrepancies between these ground measurements and the true on-orbit performance. This is not entirely unexpected, as a perfect optical system with zero distortion is physically unachievable. This also suggests the possibility of uncharacterized errors in the experimental setup of the pre-launch ground tests.

The set of intrinsic parameters to be optimized consists of seven variables: the focal length $(f)$, principal point $(x_\mathrm{c},y_\mathrm{c})$, and radial distortion coefficients $(k_2, k_4, k_6)$. Although the focal length can differ between the cross- and along-track directions, a single symmetric value was optimized for both; this simplification is justified because PolCam uses a square CCD and a circular optical system. A set of shared intrinsic parameters was co-estimated with the extrinsic parameters in Section~\ref{sec:s51}, under the initial assumption that all channels possessed identical optical characteristics. The resulting shared parameters were:
\begin{align}\label{eq:e7}
f &= 74.881416~\mathrm{mm} \\
(x_\mathrm{c}, y_\mathrm{c}) &= (512.565, 512.691)~\mathrm{pixel}\nonumber \\
(k_2, k_4, k_6) &= (6.407\times10^{-5}, 2.998\times10^{-6}, 2.928\times10^{-8})\nonumber
\end{align}
These shared values then served as the initial condition for a channel-specific estimation, yielding the intrinsic parameters for each of the six channels (summarized in Table~\ref{tab:t2}).

Figure~\ref{fig:f7} illustrates the distant errors between corresponding tie points on the image plane---measured in the cross-track, along-track, and radial directions---both before and after camera calibration. Prior to camera calibration, the PolCam-L optical system exhibits distortion of $\sim$$20~\mathrm{pixels}$ at the CCD edges [Figures~\ref{fig:f7}(a)--(c)]. By applying the refined camera parameters from Table~\ref{tab:t2}, this distortion was reduced to $\lesssim$$1.5$~pixels across the detector [Figures~\ref{fig:f7}(d)--(f)]. While the correction in the cross-track direction achieved sub-pixel precision [Figure~\ref{fig:f7}(d)], a larger residual error remains in the along-track direction [Figure~\ref{fig:f7}(e)]. This error is a direct consequence of a PolCam design feature: only six specific lines ($48$, $180$, $368$, $564$, $724$, $884$) out of $1024$ are used for observation (see Figure~\ref{fig:f1}). This sparse sampling confines the data used for optimizing the optical distortion function to a limited portion of the focal plane, which in turn limits the model's accuracy in the along-track direction.

Since the coordinate transformation for Figure~\ref{fig:f7} involves calculating the spacecraft's ephemeris with the corrected observation time and incorporating the SLDEM to remove topographic distortion, Figures~\ref{fig:f7}(d)--(f) represent the results of both the camera calibration and the complete geometric calibration. We verified the accuracy of the geometric calibration by calculating the root-mean-square error ($\mathrm{RMSE}$) of the distortion using these 160,256 geometrically-corrected tie points. The $\mathrm{RMSE}$ in the cross-track direction ($\mathrm{RMSE_{cross}}$) is $\sim$$0.87$~pixels, while the $\mathrm{RMSE}$ in the along-track direction ($\mathrm{RMSE_{along}}$) is $\sim$$1.48$~pixels. While $\mathrm{RMSE_{cross}}$ achieved sub-pixel accuracy for all channels, $\mathrm{RMSE_{along}}$ was $\sim$$1~\mathrm{pixel}$ for the central channels (2--5) and $>$$1~\mathrm{pixel}$ for the outer channels (1 and 6).

\begin{figure*}[!p]
\centering
\includegraphics[width=0.95\linewidth]{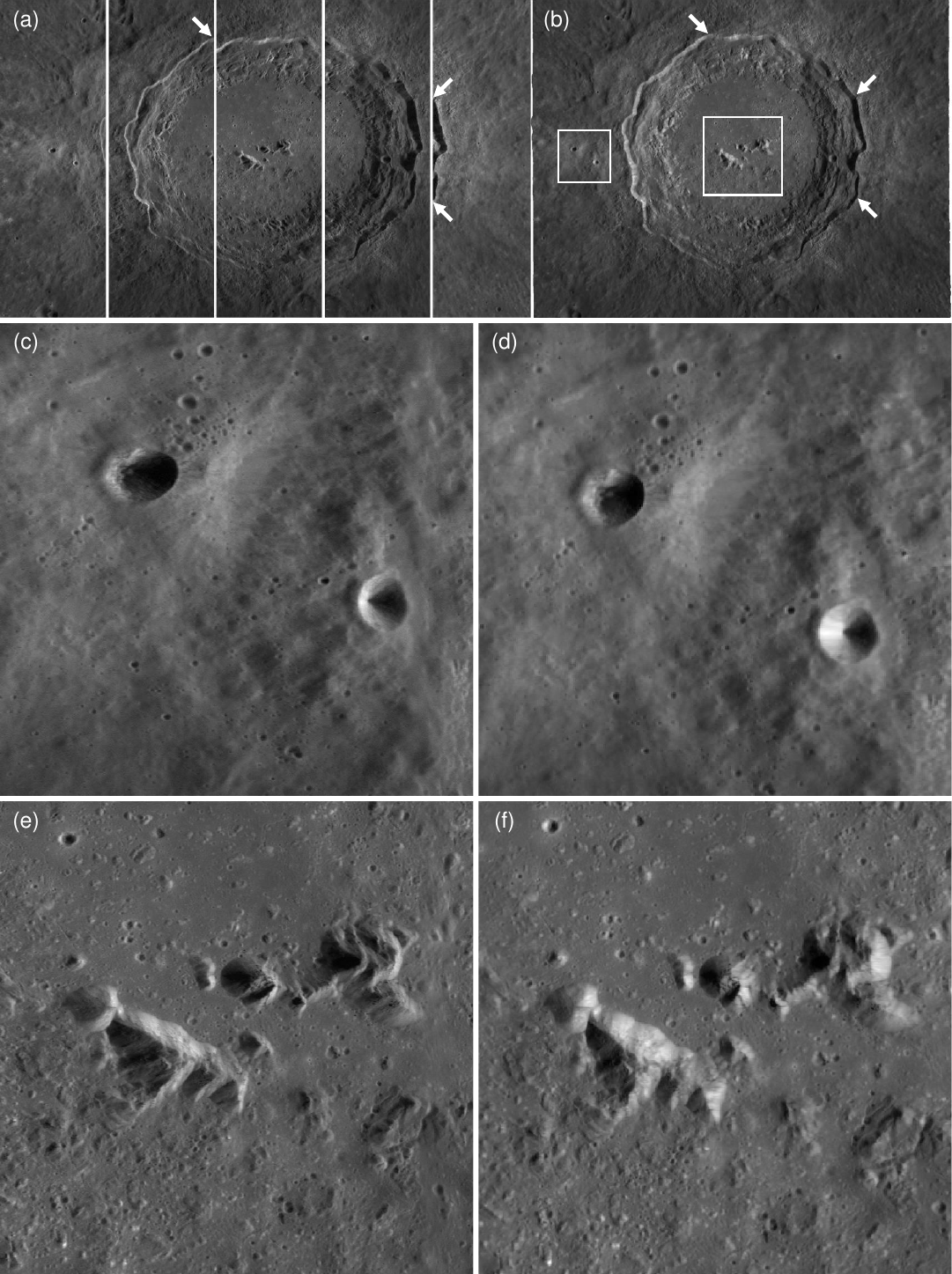}
\caption{Copernicus crater before [(a), (c), and (e)] and after [(b), (d), and (f)] geometric calibration. (a) A simple mosaic of five pre-processed PolCam observation strips covering the crater. The white arrows indicate a duplicated crater rim, highlighting significant geolocation misalignment between orbits. (b) Seamless mosaic of the same data in Panel (a) after the calibration and orthorectification. (c), (d) Magnified views of the small craters within the left white box in Panel~(b). (e), (f) Magnified views of the central peak within the right white box in Panel~(b).}\label{fig:f8}
\end{figure*}

\begin{figure*}[!t]
\centering
\includegraphics[width=\linewidth]{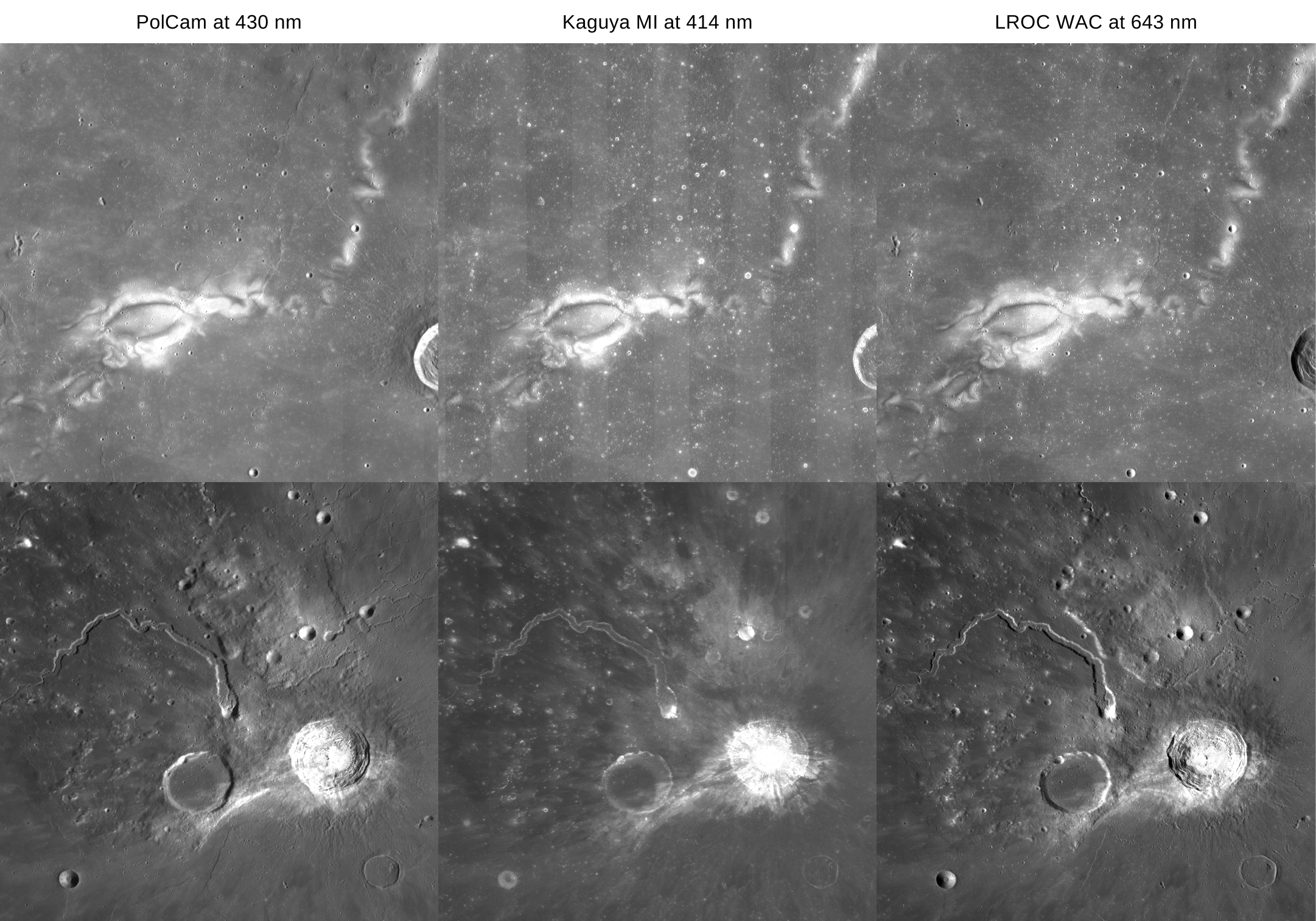}
\caption{Orthorectified and mosaicked images of Reiner Gamma (top row) and Aristarchus Plateau (bottom row) from (left) PolCam at 430 nm, (middle) Kaguya MI at 414 nm, and (right) LROC WAC at 643 nm. Each set of images for a given target covers the same latitude and longitude range.}\label{fig:f9}
\vspace{2mm}
\end{figure*}

\newpage
\section{Orthorectification}\label{sec:s6}

The final step in the geometric calibration of lunar orbiter data is orthorectification, which involves a topographically-corrected map projection into the standard $\mathrm{MOON}\_\mathrm{ME}$ coordinate system. This process accurately projects the distorted, observed imagery onto a uniform grid, an essential step for comparative analysis with lunar maps from other missions. For polarimetric analysis in particular, precise co-registration is critical---both between channels acquired simultaneously and between images acquired at different times (and thus different phase angles). Therefore, orthorectification represents the crucial step in producing geometrically-corrected data products for PolCam as well as other lunar missions.

As detailed in Sections~\ref{sec:s4} and~\ref{sec:s5}, the observation timing and camera parameters were successfully estimated using on-orbit data of PolCam, completing the preparation for estimating high-precision 3D coordinates for each pixel of an observed image. Using the \texttt{sincpt} function of the SPICE Toolkit, we compute the primary data products (latitude, longitude, and elevation) and ancillary data products (incidence, reflectance, and phase angles), all of which are generated with the same dimensions as the original observation data. These calibrated coordinate grids are then used to generate an orthorectified image, projected onto an equidistant cylindrical coordinate system. This process removes geometric distortions caused by surface topography using \texttt{patch orthorectification}, a technique common in planetary missions (see Section~\ref{sec:s3}).

To validate the orthorectified image, the final product of our geometric calibration, we selected Copernicus crater as a test site. This location is suitable as it is covered by multiple PolCam observation strips and contains numerous features susceptible to significant topographic distortion, such as steep crater walls, a central peak, and various small craters. Complete coverage of Copernicus crater and its vicinity requires five PolCam observation strips [Figure~\ref{fig:f8}(a)], with each strip having approximately 10\% cross-track overlap with the adjacent strip. Figure~\ref{fig:f8}(b) shows the seamless mosaic after these strips have been projected onto an equidistant cylindrical map. The quality of the co-registration is evident in these duplicated regions, where even topographically complex features, such as steep crater walls and rim crests with slopes exceeding $30^{\circ}$ [indicated by white arrows in Figures~\ref{fig:f8}(a) and~\ref{fig:f8}(b)] are overlapped seamlessly.

Figure~\ref{fig:f8}(c) provides a magnified view of two small craters ($\sim$$2~\mathrm{km}$ in diameter) on the west of Copernicus crater. Due to PolCam's oblique viewing angle, these features exhibit severe geometric distortion in the raw data, causing their circular shapes to appear compressed and warped. After geometric calibration, however, their true circular morphology is correctly restored [Figure~\ref{fig:f8}(d)]. Figures~\ref{fig:f8}(e) and~\ref{fig:f8}(f) show a similar comparison for the central peak of Copernicus. The pre-calibration image [Figure~\ref{fig:f8}(e)] exhibits significant foreshortening, a common artifact of oblique viewing. The orthorectified image [Figure~\ref{fig:f8}(f)] successfully corrects this distortion, restoring the peak to its proper, nadir-equivalent perspective.

Figure~\ref{fig:f9} provides a comparison between the ortho-rectified and mosaicked PolCam data and the maps from Kaguya MI and LROC WAC. The images, which are cropped to the same latitude and longitude range, demonstrate that the distinctive features of the Reiner Gamma swirl and Aristarchus Plateau are well-aligned across all datasets. We confirmed this high degree of co-registration by overlaying the images from the three different missions. Furthermore, other selected regions that were processed with priority, such as irregular mare patches, Marius Hills, and Rumker Peak, also showed strong alignment, validating the robustness of our method. Additionally, the PolCam images in the left column of \mbox{Figure~\ref{fig:f9}} are \mbox{unpolarized} $430~\mathrm{nm}$ images created by combining \mbox{channels~\mbox{2--4}}, which also confirms the excellent inter-channel \mbox{alignment}.

\section{Summary and Discussions}\label{sec:s7}

The primary objective of this study was to estimate observation timing and refine the camera parameters of PolCam via on-orbit geometric calibration, with the ultimate goal of using these refined parameters to generate well-orthorectified PolCam data. To establish the reference ground truth, we constructed a network of tie points by detecting and matching bright features on the PolCam images and the well-orthorectified Kaguya MI map (Figure~\ref{fig:f2}). This network of tie points was instrumental in resolving two critical issues of PolCam:
\begin{enumerate}
\item[(1)] Observation timing (see Section~\ref{sec:s4})\\
PolCam cannot determine its observation times autonomously, as it has no an independent internal clock and does not record this information in its telemetry headers. Therefore, the observation time for each horizontal line of an observation strip must be reconstructed based on the imaging start time from the spacecraft clock (with which PolCam synchronizes upon power-on) and the instrument's internal scan rate. We demonstrate that after precisely estimating the imaging start time and the optimal scan rate, the timing of each line can be accurately computed by a simple linear model [Equation~\eqref{eq:e6}]. The imaging start time for each orbit must be corrected for the offset between the spacecraft clock and standard time, which can be as large as $\pm0.4~\mathrm{sec}$ (Figure~\ref{fig:f5}). The optimal scan rate was estimated to be $26.978~\mathrm{msec}$, a value slightly shorter than the commanded rate of $27.050~\mathrm{msec}$ (Figure~\ref{fig:f6}).
\item[(2)] Camera parameters (see Section~\ref{sec:s5})\\
PolCam's extrinsic parameters (i.e., its mounting \mbox{angles}) were not measured during pre-flight testing, and the pre-flight measurements of its intrinsic parameters (focal length, principal point, and optical distortion coefficients) were found to be inaccurate for precise calibration. Therefore, by minimizing the reprojection error, defined as the Euclidean distance between matched tie points on the image plane, we first derived a common set of extrinsic parameters for the PolCam instrument. Subsequently, we determined the intrinsic parameters for each individual channel of PolCam-L camera. The resulting parameters, summarized in Table~\ref{tab:t2}, have sufficient accuracy to enable the precise camera calibration of PolCam data (Figure~\ref{fig:f7}).
\end{enumerate}

Despite the aforementioned initial limitations, this study represents a significant achievement in on-orbit geometric calibration. By relying exclusively on PolCam images acquired in orbit and the Kaguya MI map, we have successfully compensated for the limited pre-launch preparations. The methods detailed in this paper have achieved a geometric precision comparable to that of global maps from previous lunar orbiters such as Kaguya MI and LRO WAC (Figure~\ref{fig:f9}). This work completes the foundational steps necessary to produce geometrically-corrected data products of PolCam (see Section~\ref{sec:s6}).

However, the larger residual error in the along-track direction, a limitation caused by PolCam's sparse sampling (using only~$6$ of $1024$~lines), needs to be further improved. While channels $2$--$5$ achieve sub-pixel accuracy, channels~$1$ and~$6$, which are farthest from the principal point, retain a residual error of $\sim$$1.5~\mathrm{pixels}$ at the CCD edge [Figure~\ref{fig:f7}(e)]. These outer channels exhibit greater image distortion, which results in fewer tie points being detected. The smaller number of tie points for these channels consequently leads to a larger reprojection error.

The present study utilized only the data acquired during Danuri's nominal mission phase, which was conducted from a $100~\mathrm{km}$ altitude and concluded on February 18, 2025. The spacecraft has since transitioned to an extended mission at a lower $60~\mathrm{km}$ altitude. As of August 2025, we have acquired about six months of this higher-resolution data ($\sim$$20~\mathrm{m/pixel}$, an improvement from $\sim$$43~\mathrm{m/pixel}$). We speculate that repeating our calibration adding this higher-resolution dataset will provide more tie points for the distortion model, subsequently reducing the residual error in the along-track direction. This analysis is a planned focus of our future work.


\acknowledgments
The work by SSK was supported by the Korea Astronomy and Space Science Institute under the R\&D program (2023-1-850-09) supervised by the Ministry of Science and ICT.



\balance
\bibliography{PolCamGeoCal}

\end{document}